\documentclass[journal,twoside]{IEEEtran}

\usepackage[T1]{fontenc}
\usepackage[utf8]{inputenc}
\usepackage{enumerate}

\usepackage{cite}

\usepackage{tikz}
\usetikzlibrary{snakes}

\usepackage[tbtags,cmex10]{amsmath}
\usepackage{amssymb}
\usepackage{amsthm}
\usepackage{amsfonts}

\usepackage{subcaption}

\usepackage{float}
\usepackage{fixltx2e}

\newcommand{\field}[1][]{\ensuremath{\mathbb{F}_{#1}}}
\newcommand{\code}{\ensuremath{\mathcal{C}}}
\newcommand{\floor}[1]{\left\lfloor #1 \right\rfloor}

\newcommand{\weightdist}[2]{A^{#1}_{#2}}
\newcommand{\wordsaround}[2]{N^{#1}_{#2}}
\newcommand{\decoderchanges}[2]{D^{#1}_{#2}}

\theoremstyle{plain}
\newtheorem{thm}{Theorem}
\newtheorem*{thm*}{Theorem}
\newtheorem{prop}[thm]{Proposition}
\newtheorem{cor}[thm]{Corollary}
\newtheorem*{cor*}{Corollary}
\newtheorem{lem}[thm]{Lemma}

\theoremstyle{remark}
\newtheorem{rem}[thm]{Remark}

\theoremstyle{definition}

\begin{document}


\title{Information--bit Error Rate and False Positives \\ in an MDS Code}

\author{F. J. Lobillo, Gabriel Navarro, and Jos\'e G\'omez-Torrecillas%
\thanks{Research partially supported by grant MTM2010-20940-C02-01 from the 
Ministerio de Ciencia e Innovaci\'on of the Spanish Government and from FEDER.}%
\thanks{F. J. Lobillo is with the Department of Algebra, ETSI Inform\'atica y de Telecomunicaci\'on, University of Granada, email: jlobillo@ugr.es}%
\thanks{Gabriel Navarro is with the Department of Computer Science and Artificial Intelligence, University of Granada, email: gnavarro@ugr.es}%
\thanks{Jos\'e G\'omez-Torrecillas is with the Department of Algebra, Facultad de Ciencias, University of Granada, email: gomezj@ugr.es}}

\markboth{Information--bit error rate and false positives in an MDS code}%
{Lobillo, Navarro and G\'omez-Torrecillas}

\maketitle

\begin{abstract}
In this paper, a computation of the input-redundancy weight enumerator is presented. This is used to improve the theoretical approximation of the information--bit and --symbol error rate, in terms of the channel bit-error rate, in a block transmission through a discrete memoryless channel. Since a bounded distance reproducing encoder is assumed, the computation of the here-called false positive (a decoding failure with no information-symbol error) is provided. As a consequence, a new performance analysis of an MDS code is proposed.
\end{abstract}


\section{Introduction}

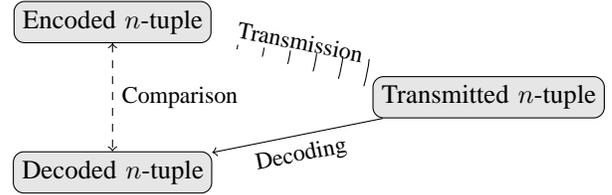
\begin{figure}[!ht]
\begin{displaymath}\begin{tikzpicture}
\draw (0,1) node(uno) [draw, shape=rectangle, rounded corners, fill=gray!20] {Encoded $n$-tuple};
\draw (5,0) node(dos) [draw, shape=rectangle, rounded corners, fill=gray!20] {Transmitted $n$-tuple};
\draw (0,-1) node(tres) [draw, shape=rectangle, rounded corners, fill=gray!20] {Decoded $n$-tuple};
\draw[decorate,decoration={expanding waves,angle=5}] (uno) to  node[snake=snake, midway, sloped, above, pos=0.5] {\small Transmission} (dos) ;
\draw [->] (dos) to  node[midway, sloped, below, pos=0.5] {\small Decoding} (tres);
\draw [<->, dashed] (uno) to node[midway, right] {\small Comparison} (tres);
\end{tikzpicture}\end{displaymath}
\caption{Length oriented encoding-decoding scheme.}\label{f1}
\end{figure}

\IEEEPARstart{A}{fundamental} challenge when determining the performance of a block error-correcting code is to measure its bit-error rate (BER), which quantifies the reliability of the system. In practice, the BER estimation for a single code is simple, just send data and divide the errors committed among the total number of bits. However, it would be too costly and time-consuming if a comparative between several codes is required. Mathematical software packages for encoding and decoding are very limited and restricted to specific codes and simulations would consume a huge amount of time when dealing with low bit-error rates. For this reason, a theoretical approach to the measurement of the BER is proposed by several authors in the literature, see for instance \cite{KasamiLin1984, Dodunekova1997, Dodunekova2002, DessetMacqVandendorpe2004, Han2006, Dodunekova2008, Yar2008}. 
All these papers follow this scheme of codification: let \code{} be a code of length $n$ and dimension $k$ over the field with $q$ elements, being $q\geq 2$. An $n$-tuple is transmitted through a $q$-ary symmetric discrete memoryless channel. In this step, there are two possibilities, the transmission is right or failed in some symbols. In a second step, the code corrects the $n$-tuple, detects an erroneous transmission but does not correct it, or asserts that the transmitted $n$-tuple is a codeword. Finally, there is a comparison between the encoded and decoded $n$-tuples, see Figure \ref{f1}.
When we run over all the possibilities in each step (of course, not all combinations are possible), this yields five disjoint cases:
\begin{enumerate}
\item A correct transmission (CT), i.e., all the symbols are correctly received.
\item A right correction (RC), i.e., some of the symbols are incorrectly received but the decoding  algorithm corrects them.
\item An error detection (ED), i.e., the number of errors exceeds the error-correction capability of the code, the block is not corrected and the bad transmission is detected.
\item A wrong correction (WC), i.e., some errors occur (beyond the error capability of the code), there is a code-correction, but nevertheless, the encoded block differs from the decoded block.
\item A false negative (FN), i.e., some symbols are incorrectly received but the whole block is a codeword, so, from the receiver's point of view, the block is correctly received. 
\end{enumerate}

Cases FN and WC are called undetected errors in \cite{KasamiLin1984}, and it is proven that, for maximum-distance-separable (MDS) codes, the probability of an undetected error decreases monotonically as the channel symbol error decreases, that is, MDS codes are proper in the terminology of \cite{Dodunekova1997}. Hence, the performance of an MDS code
is characterized by the probability that an erroneous transmission will remain undetected. In \cite{DessetMacqVandendorpe2004}, as a performance criterion, the probability of an ED is added and an exhaustive calculus of the word--, symbol--, and bit--error rate of ED, WC, and FN is made.

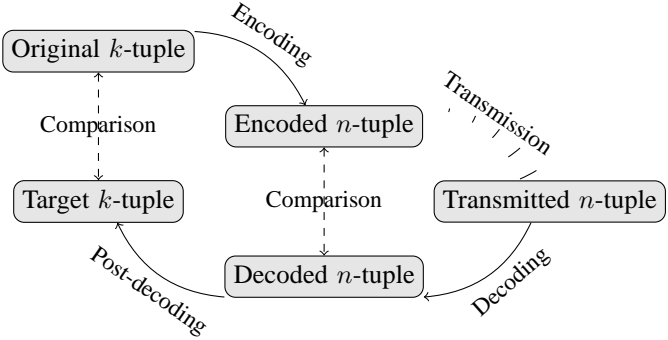
\begin{figure}[!ht]
\begin{displaymath}\begin{tikzpicture}
\draw (-3,2) node(cero) [draw, shape=rectangle, rounded corners, fill=gray!20] {Original $k$-tuple};
\draw (0,1) node(uno) [draw, shape=rectangle, rounded corners, fill=gray!20] {Encoded $n$-tuple};
\draw (3,0) node(dos) [draw, shape=rectangle, rounded corners, fill=gray!20] {Transmitted $n$-tuple};
\draw (0,-1) node(tres) [draw, shape=rectangle, rounded corners, fill=gray!20] {Decoded $n$-tuple};
\draw (-3,0) node(cuatro) [draw, shape=rectangle, rounded corners, fill=gray!20] {Target $k$-tuple};
\draw [decorate,decoration={expanding waves,angle=4}] (uno) to [bend left] node[midway, sloped, above, pos=0.5] {\small Transmission} (dos);
\draw [->] (dos) to [bend left] node[midway, sloped, below, pos=0.4] {\small Decoding} (tres);
\draw [<->, dashed] (uno) to node[midway] {\small Comparison} (tres);
\draw [->] (cero)  to[bend left] node[midway, sloped, above] {\small Encoding} (uno) ;
\draw [->] (tres) to[bend left] node[midway, sloped, below] {\small Post-decoding}
(cuatro);
\draw [<->, dashed] (cero) to node[midway] {\small Comparison} (cuatro);
\end{tikzpicture}\end{displaymath}
\caption{Dimension oriented encoding-decoding scheme.}\label{f2}
\end{figure}

In this paper, we propose a refinement in the calculi of the probability of an FN, a WC, and an ED. Consequently, we get a better approximation of the BER for a $q$-ary MDS code. As in the above references, we consider a bounded distance
reproducing decoder, i.e., it reproduces the received word whenever there are uncorrectable errors. 
The underlying idea consists in removing the symbol errors produced in the redundancy part of the decoded $n$-tuple, that is, following the nomenclature of \cite{Torrieri1984, Torrieri1988}, unlike the aforementioned papers, we estimate the information--bit error rate (iBER), sometimes also called post-decoding bit--error rate. More formally, let us assume, without loss of generality, that the codification is systematic and the first $k$ symbols of the $n$-tuples form an information set. Hence, following the above scheme, after the comparison step, if there are $s$ errors, the symbol-error proportion is $s/n$. Nevertheless, some of these errors belong to the redundancy part and they will not spread in the final post-decoded $k$-tuple. In other words, a new variable should be considered: the comparison between the original block and the final $k$-tuple obtained after decoding, see Figure \ref{f2}.

Attending to this new variable, we may split the ED into two disjoint cases:
\begin{enumerate}[{3}.a)]
\item A pure error detection (PED), i.e., some errors affect the information set.
\item A false positive (FP), i.e., all errors belong to the redundancy part. So, from the point of view of the receiver, there are uncorrectable errors but, indeed, the post-decoded block is right.
\end{enumerate}

Then, a study of the BER should consider the probability of obtaining a false positive after the post-decoding process and, hence, the criterion for measuring the performance of the code should be the probability of undetected and PED errors.

All along this paper we shall assume that the code under consideration is a $q$-ary $[n,k]$ MDS  code in a bounded distance reproducing encoder, where $q=2^b$ for some natural $b>0$ 
 and $0<k<n$, with minimum distance $d=n-k+1$ and error-correcting capacity up to $t=\floor{(n-k)/2}$ errors. Furthermore, we shall assume, without loss of generality, that the generator matrix is systematic in its first $k$ coordinates. This is so since it is well known that, in an MDS code, any subset of $k$ coordinates form an information set, see e.g. \cite[Theorem 5.3.4]{Roman1992}. The reorganization of these components to the first $k$ does not affect our calculations and makes the text much more readable. The channel BER shall be denoted by $p$, and by $1-p$, the probability of a bit of being correctly received. The code is $2^b$-ary, so the probability for a symbol of being correctly received is $q_s=(1-p)^b$. Therefore, the expected probability of transition between any two symbols, the channel symbol-error rate, is the average
of all possible cases, i.e., $p_s=\frac{1-(1-p)^b}{q-1}$. Finally, if we
want to know the probability for a bit to be erroneous inside an
erroneous symbol, we simply give the conditional probability $p_{b|s}=\frac{p}{1-(1-p)^b}$.
We will denote by $\wedge$ and $\vee$ the minimum and the maximum of a set of integers.

The paper is structured as follows. In Section \ref{s2}, we count the codewords of a $q$-ary MDS code. The number of codewords of a fixed weight has been calculated before (see \cite{KasamiLin1984} and \cite{Macwilliams1963}); nevertheless, we need a finer approach. In \cite{KhamyMcEllice2005} and \cite{Khamy2007} a formula for the number of codewords of fixed weight with respect to a given partition is provided. By means of arguments of Linear Algebra, we shall give an equivalent formula for counting the number of codewords of a fixed weight both in the information set and in the redundancy part. This allows us to calculate the iBER of an FN. In Section \ref{s3}, we shall deal with the decoding failures, that is, when the received word is corrected to a different codeword to the original one. We shall follow the style of \cite{DessetMacqVandendorpe2004} and calculate the words inside the sphere of a codeword of a given  information and redundancy weight. We will make use of these calculations in Section \ref{s4} in order to give an expression of the probability that an FP occurs and obtain the desired approximation of the BER of an MDS code. In order to make the paper self-contained, in the Appendix, we add some of the combinatorial formulae needed all along the text.

Figures \ref{graphFN} 
to \ref{graphPED0} have been made with \textsf{Sage} \cite{sage}.

\section{Counting codewords}\label{s2}

Let \code{} be a $q$-ary $[n,k]$ MDS code generated by a systematic matrix $G=(I_k|R)$. The aim of this section is to compute $\weightdist{i}{j}$, the number of codewords in \code{} with $i$ non-zero information symbols and $j$ non-zero redundancy symbols, where $i\in\{0,\cdots , k\}$ and $j\in\{0, \cdots, n-k\}$. This is called the \emph{input-redundancy weight enumerator} (IRWE) in the literature (see e.g. \cite{KhamyMcEllice2005}). In fact the weight enumerator of any partition $\mathcal{T}$ on $\{1, \dots, n\}$ is computed in \cite[Theorem 1]{KhamyMcEllice2005} and \cite[Theorem 3.1]{Khamy2007}, hence $\weightdist{i}{j}$ can be obtained as a particular case of those theorems. We propose a new way to compute $\weightdist{i}{j}$ involving linear algebra techniques. We shall need the following lemmata.

\begin{lem}\label{matrixMDScfr}
Any square submatrix of $R$ is regular.
\end{lem}

\begin{IEEEproof}
Let 
\begin{displaymath}
R = \begin{pmatrix}
r_{11} & \dots & r_{1,n-k} \\
\vdots & \ddots & \vdots \\
r_{k1} & \dots & r_{k,n-k}
\end{pmatrix}
\end{displaymath}
and let $\rho = \{i_1, \dots, i_l\}$ and $\gamma = \{j_1, \dots, j_l\}$ be the sets of indices corresponding to the rows and columns of $R$ that form an $l \times l$ submatrix, where $l\leq k\wedge (n-k)$. We call such submatrix $R_{\rho\gamma}$. If $\det(R_{\rho\gamma}) = 0$, there exists a non-zero vector $(a_1, \dots, a_l) \in \field[q]^l$ such that
\begin{equation}\label{eq:Rrc-singular}
(a_1 \dots a_l)
\begin{pmatrix}
r_{i_1j_1} & r_{i_1 j_2} & \cdots & r_{i_1j_l} \\
r_{i_2j_1} & r_{i_2 j_2} & \cdots & r_{i_2j_l} \\
\vdots & \vdots & \ddots & \vdots \\
r_{i_lj_1} & r_{i_l j_2} & \cdots & r_{i_lj_l}
\end{pmatrix}
= (0 \dots 0).
\end{equation}
Let $a \in \field[q]^k$ be the vector in which $a_h$ is in the $i_h$-th coordinate for all $1 \leq h \leq l$ and zero otherwise. Since $aG = (aI_k | aR)$, there are $l$ non-zero coordinates in the first $k$ coordinates of $aG$. Now, for any $1 \leq h \leq l$
\begin{displaymath}
a \begin{pmatrix}
r_{1j_h} \\
r_{2 j_h} \\
\vdots \\
r_{k j_h}
\end{pmatrix} =
a_1 r_{i_1j_h} + a_2 r_{i_2 j_h} + \dots a_l r_{i_l j_h} = 0
\end{displaymath}
by \eqref{eq:Rrc-singular}, so, in the last $n-k$ coordinates of $aG$, there are no more than $n-k-l$ non-zero coordinates. Then, $aG$ has weight $n-k$ at most. Since $aG \in \code$ and the minimum distance of \code{} is $d=n-k+1$, we get a contradiction. Consequently, $R_{\rho\gamma}$ is regular.
\end{IEEEproof}

Given a system of linear equations, we say that a solution is totally non-zero if all the coordinates of the vector are other than zero. We say that a matrix has totally full rank if any submatrix has full rank. By Lemma \ref{matrixMDScfr}, $R$ has totally full rank.

Let us denote by $f^q_{i,j}$ the number of totally non-zero solutions of a homogeneous linear system,  over the field \field[q], with $i$ variables and $j$ equations whose coefficient matrix has totally full rank.

\begin{lem}\label{fqijrecursiva}
For any integers $i, j\geq 1$,  $f^q_{i,j}$ is given by the following recurrence:
\begin{equation}\label{eq:recurSolNoNulas}
f^q_{i,j}=\begin{cases}
0 & \text{if $i\leq j$}\\
(q-1)^{i-j}-\sum_{h=1}^j \binom{j}{h} f^q_{i-h,j} & \text{if $i>j$}
\end{cases}
\end{equation}
\end{lem}

\begin{IEEEproof}
If $i\leq j$, there are, at least, as many equations as variables. Since its coefficient matrix has full rank, the zero vector is the only solution. Then $f^q_{i,j}=0$.

If $i>j$, since its coefficient matrix has full rank, the system is undetermined. Specifically, whenever we fix $i-j$ coordinates, we will find a single unique solution. Then, there are $(q-1)^{i-j}$ solutions whose first $i-j$ coordinates are non-zero. In order to calculate $f^q_{i,j}$, it is enough to subtract those solutions for which some of the remaining $j$ coordinates are zero. For any $0<h\leq j$, the solutions with exactly $h$ coordinates being zero may be obtained choosing $h$ coordinates from $j$, and calculating the number of totally non-zero solutions in a linear system of $i-h$ variables and $j$ equations, that is, $\binom{j}{h}f^q_{i-h,j}$. Therefore,
\begin{displaymath}
f^q_{i,j}=(q-1)^{i-j}-\sum_{h=1}^j \binom{j}{h} f^q_{i-h,j}.
\end{displaymath}
\end{IEEEproof}

\begin{prop}\label{fqijdirecta}
For any integers $i,j\geq 1$,
\begin{displaymath}
f^q_{i,j}=\sum_{l=0}^{i-j-1}(-1)^l \binom{j+l-1}{l}(q-1)^{i-j-l}
\end{displaymath}
\end{prop}

\begin{IEEEproof}
By Lemma \ref{fqijrecursiva}, it is enough to check that the expression $F^q_{i,j} = \sum_{l=0}^{i-j-1}(-1)^l \binom{j+l-1}{l}(q-1)^{i-j-l}$ satisfies \eqref{eq:recurSolNoNulas}. Observe that if $i \leq j$, then $F^q_{i,j}=0$, since the sum runs over the empty set. Suppose that $i>j$.
Since $F^q_{i-h,j} = 0$ for any $h \geq i-j$, we obtain 
\begin{equation}\label{eq:recF}
\sum_{h=0}^j \binom{j}{h} F^q_{i-h,j} = \sum_{h=0}^{j \wedge (i-j-1)} \binom{j}{h} F^q_{i-h,j}.
\end{equation}
We substitute $f^q_{i-h,j}$ by $F^q_{i-h,j}$ in \eqref{eq:recurSolNoNulas}, and using \eqref{eq:recF}, we get
\begin{displaymath}
\begin{split}
&\sum_{h=0}^{j \wedge (i-j-1)} \binom{j}{h} F^q_{i-h,j} = \\
&= \sum_{h=0}^{j \wedge (i-j-1)} \binom{j}{h} \sum_{l=0}^{i-h-j-1} (-1)^l \binom{j+l-1}{l}(q-1)^{i-h-j-l} \\
&= \sum_{h=0}^{j \wedge (i-j-1)} \sum_{l=0}^{i-h-j-1} (-1)^l \binom{j}{h} \binom{j+l-1}{l}(q-1)^{i-h-j-l}.
\end{split}
\end{displaymath}
This expression is a polynomial in $(q-1)$. Let us now calculate the coefficients of $(q-1)^{i-j-H}$ for $H=0,\cdots, j\wedge (i-j-1)$. 
We group those coefficients in which $l+h=H$. If $H = 0$, then $l=h=0$, and therefore the coefficient of $(q-1)^{i-j}$ is 
\begin{displaymath}
(-1)^0 \binom{j}{0} \binom{j-1}{0} = 1.
\end{displaymath}
Suppose that $0 < H \leq j \wedge (i-j-1)$. The coefficient of $(q-1)^{i-j-H}$ is given by
\begin{displaymath}
\begin{split}
&\sum_{h=0}^H (-1)^{H-h} \binom{j}{h} \binom{j+H-h-1}{H-h} = \\ 
&= \sum_{h=0}^H (-1)^{H-h} \frac{j!}{h!(j-h)!} \frac{(j+H-h-1)!}{(H-h)! (j-1)!} \\
&= \frac{j}{H!} \sum_{h=0}^H (-1)^{H-h} \binom{H}{h} \cdot \\
&\quad \cdot (j+H-h-1) (j+H-h-2) \cdots (j-h+1) \\
&= 0,
\end{split}
\end{displaymath}
where the last equality comes from Lemma \ref{DifFin}, since $(j+H-h-1) (j+H-h-2) \cdots (j-h+1)$ is a polynomial in $h$ of degree $H-1$. 

Because of $l \leq i-h-j-1$, then $H \leq i-j-1$. As a result the case when $j < H \leq i-j-1$ is left to be analyzed. In such case, the coefficient of $(q-1)^{i-j-H}$ is given by
\begin{displaymath}
\begin{split}
& \sum_{h=0}^j (-1)^{H-h} \binom{j}{h} \binom{j+H-h-1}{H-h} = \\
&= \sum_{h=0}^j (-1)^{H-h} \binom{j}{h} \frac{(j+H-h-1)!}{(j-1)! (H-h)!} \\
&= \frac{(-1)^{H-j}}{(j-1)!} \sum_{h=0}^j (-1)^{j-h} \binom{j}{h} \cdot \\
&\quad \cdot (j+H-h-1) (j-H-h-2) \cdots (H-h+1) \\
&= 0,
\end{split}
\end{displaymath}
where, again, the last equality follows from Lemma \ref{DifFin} since $(j+H-h-1) (j-H-h-2) \cdots (H-h+1)$ is a polynomial in $h$ of degree $j-1$. Then, 
\begin{displaymath}
\sum_{h=0}^{j\wedge i-j-1} \binom{j}{h} F^q_{i-h,j} = (q-1)^{i-j}.
\end{displaymath}
By \eqref{eq:recF},
\begin{displaymath}
F^q_{ij}=(q-1)^{i-j}-\sum_{h=1}^{j} \binom{j}{h} F^q_{i-h,j},
\end{displaymath}
so $F^q_{i,j} = f^q_{i,j}$. 
\end{IEEEproof}

We recall that $\weightdist{i}{j}$ is the number of codewords of \code{} with weight $i$ in the first $k$ coordinates (the information set) and $j$, in the remaining $n-k$ coordinates (the redundancy part). 

\begin{lem}\label{Ankijviafqij}
Let $0 \leq j \leq n-k$. If $1 \leq i \leq k$, then
\begin{displaymath}
\weightdist{i}{j} = \binom{k}{i} \binom{n-k}{j} \sum_{l=0}^j (-1)^l \binom{j}{l} f^q_{i,n-k-j+l}. 
\end{displaymath}
If $i=0$, then $\weightdist{i}{j}=\delta_{ij}$, where $\delta_{ij}$ is the Kroneker's delta. 
\end{lem}

\begin{IEEEproof}
The case $i=0$ is evident because the minimum distance of \code{} is $d=n-k+1$. Suppose that $i \geq 1$. A codeword with weight $i$ in the information set and $j$ in the redundancy part must be obtained as a linear combination of $i$ rows of $G$ with non zero coefficients such that exactly $n-k-j$ coordinates of the redundancy part become zero, i.e. it is a totally non-zero solution of a homogeneous linear system with $n-k-j$ equation and $i$ variables. By Lemma \ref{matrixMDScfr}, this number is given by $f^q_{i,n-k-j}$. However some of the solutions counted in $f^q_{i,n-k-j}$ can turn zero some of the remaining $j$ coordinates. Hence, the lemma is obtained as a consequence of the Inclusion--Exclusion Principle.
\end{IEEEproof}

\begin{rem}\label{Ankij=0}
Observe that if $i+j \leq n-k$, then $i \leq n-k-j+l$, so $\weightdist{i}{j} = 0$. This agrees with the fact that the minimum weight of a non-zero codeword must be $n-k+1=d$.
\end{rem}

Let us now give a neater description of $\weightdist{i}{j}$. By mixing Proposition \ref{fqijdirecta} and Lemma \ref{Ankijviafqij}, we have
\begin{equation}\label{eq:Ankij}
\begin{split}
&\weightdist{i}{j} = \\
&= \binom{k}{i} \binom{n-k}{j} \sum_{l=0}^j (-1)^l \binom{j}{l} \cdot \\
&\quad \cdot \sum_{h=0}^{i-(n-k-j+l)-1} (-1)^h \binom{n-k-j+l+h-1}{h} \cdot \\
&\quad \cdot (q-1)^{i-(n-k-j+l)-h} \\
&= \sum_{l=0}^j \sum_{h=0}^{i-n+k+j-l-1} (-1)^{l+h} \cdot \\
&\quad \cdot \binom{k}{i} \binom{n-k}{j} \binom{j}{l} \binom{n-k-j+l+h-1}{h} \cdot \\
&\quad \cdot (q-1)^{i-n+k+j-l-h}.
\end{split}
\end{equation}

We proceed analogously to the proof of Proposition \ref{fqijdirecta}. Let us examine the coefficient of $(q-1)^{i-n+k+j-H}$ in \eqref{eq:Ankij}, where $l+h=H$. Since $h \leq i-n+k+j-l-1$, it is necessary that $H \leq i-n+k+j-1$. We distinguish two cases. On the one hand, if $H \leq j$, then $0 \leq H \leq j \wedge (i-n+k+j-1)$, and then, the coefficient is given by
\begin{equation}\label{eq:Ankij(inkjH)H<j}
\begin{split}
&\sum_{l=0}^H (-1)^H \binom{k}{i} \binom{n-k}{j} \binom{j}{l} \binom{n-k-j+H-1}{H-l} = \\
&= (-1)^H \binom{k}{i} \binom{n-k}{j} \sum_{l=0}^H \binom{j}{l} \binom{n-k-j+H-1}{H-l} \\
&= (-1)^H \binom{k}{i} \binom{n-k}{j} \binom{n-k+H-1}{H},
\end{split}
\end{equation}
where the last equality is a consequence of Lemma \ref{ChuVandermonde}. On the other hand, if $j \leq H$, the coefficient of $(q-1)^{i-n+k+j-H}$ is given by 
\begin{equation}\label{eq:Ankij(inkjH)j<H}
\begin{split}
&\sum_{l=0}^j (-1)^H \binom{k}{i} \binom{n-k}{j} \binom{j}{l} \binom{n-k-j+H-1}{H-l} = \\
&= (-1)^H \binom{k}{i} \binom{n-k}{j} \sum_{l=0}^j \binom{j}{l} \binom{n-k-j+H-1}{H-l} \\
&= (-1)^H \binom{k}{i} \binom{n-k}{j} \sum_{l=0}^j \frac{j!}{l! (j-l)!} \cdot \\
&\quad \cdot \frac{(n-k-j+H-1)!}{(H-l)!(n-k-j+l-1)!} \\
&= (-1)^H \binom{k}{i} \binom{n-k}{j} \cdot \\
&\quad \cdot \sum_{l=0}^j \frac{j! (n-k-j+H-1) \cdots (n-k) (n-k-1)!}{l! (H-l)! (j-l)!(n-k-j+l-1)!} \\
&= (-1)^H \binom{k}{i} \binom{n-k}{j} \frac{(n-k-j+H-1) \cdots (n-k)}{H (H-1) \cdots (j+1)} \cdot \\
&\quad \cdot \sum_{l=0}^j \frac{H!}{l!(H-l)!} \frac{(n-k-1)!}{(j-l)!(n-k-j+l-1)!} \\
&= (-1)^H \binom{k}{i} \binom{n-k}{j} \frac{(n-k-j+H-1) \cdots (n-k)}{H (H-1) \cdots (j+1)} \cdot \\
&\quad \cdot \sum_{l=0}^j \binom{H}{l} \binom{n-k-1}{j-l} \\
&\overset{\dagger}{=} (-1)^H \frac{n-k}{H} \binom{k}{i} \binom{n-k-j+H-1}{H-1} \binom{n-k+H-1}{j} \\
&\overset{\ddagger}{=} (-1)^H \frac{n-k}{H} \binom{k}{i} \binom{n-k+H-1}{H-1} \binom{n-k}{j} \\
&= (-1)^H \binom{k}{i} \binom{n-k}{j} \frac{(n-k) (n-k+H-1)!}{H (H-1)! (n-k)!} \\
&= (-1)^H \binom{k}{i} \binom{n-k}{j} \frac{(n-k+H-1)!}{H! (n-k-1)!} \\
&= (-1)^H \binom{k}{i} \binom{n-k}{j} \binom{n-k+H-1}{H},
\end{split}
\end{equation}
where $\dagger$ is given by Lemma \ref{ChuVandermonde} and Lemma \ref{extBinom}, and $\ddagger$ is a consequence of Lemma \ref{prodBinom}. That is, by \eqref{eq:Ankij(inkjH)H<j} and \eqref{eq:Ankij(inkjH)j<H}, we have proven the main result of this section:

\begin{thm}\label{Ankij}
\begin{displaymath}
\begin{split}
&\weightdist{i}{j} = \binom{k}{i} \binom{n-k}{j} \cdot \\
&\quad \cdot \sum_{H=0}^{i-n+k+j-1} (-1)^H \binom{n-k+H-1}{H} (q-1)^{i-n+k+j-H}.
\end{split}
\end{displaymath}
\end{thm}

Recall from \cite[Theorem 1]{KhamyMcEllice2005} and \cite[Theorem 3.1]{Khamy2007} that for any $(n_1, \dots, n_p)$ partition $\mathcal{T}$ on $\{1, \dots, n\}$, the partition weight enumerator is
\begin{equation}\label{eq:pwe}
\begin{split}
&A^\mathcal{T}(w_1, \dots, w_p) = \\
&= \binom{n_1}{w_1} \cdots \binom{n_p}{w_p} \sum_{j_1=0}^{w_1} \binom{w_1}{j_1} (-1)^{w_1-j_1} 
\dots \\
&\quad \cdot \sum_{j_p=d-\sum_{z=1}^{p-1}j_z}^{w_p} \binom{w_p}{j_p} (-1)^{w_p-j_p} (q^{k-n+\sum_{z=1}^{p} j_z} - 1)
\end{split}
\end{equation}

\begin{prop}\label{pwe=Ankij}
Let $\mathcal{T}$ be the $(k,n-k)$ partition on $\{1, \dots, n\}$ associated to the $k$ information symbols and $n-k$ redundancy symbols. Then $A^\mathcal{T}(i,j) = \weightdist{i}{j}$ for all $i \leq k$ and $j \leq n-k$.
\end{prop}

The proof of Proposition \ref{pwe=Ankij} is in Appendix \ref{pf:pwe=Ankij}.

As a consequence of Theorem \ref{Ankij}, we may recover the well known formula of the weight distribution of an MDS code as it appears in \cite{KasamiLin1984}. Let $\{ A_r ~|~ 0 \leq r \leq n \}$ denotes the weight distribution of \code{}. Then 
\begin{displaymath}
A_r = \sum_{i=0}^{r \wedge k} \weightdist{i}{r-i},
\end{displaymath}
since a codeword of weight $r$ distributes its weight between the first $k$ and the last $n-k$ coordinates.

\begin{prop}\label{MDSwd}
The weight distribution of an MDS code is given by the following formula:
\begin{displaymath}
A_r = \binom{n}{r} \left[ \sum_{j=0}^{r-d} (-1)^j \binom{r}{j} q^{r-j+1-d} + \sum_{j=r-d+1}^{r} (-1)^j \binom{r}{j} \right]
\end{displaymath}
for $d \leq r \leq n$, $A_0=1$ and $A_r=0$ if $0 < r < d$.
\end{prop}

\begin{IEEEproof}
The case $r=0$ is evident. If $0 < r \leq n-k < d$, as we have pointed out in Remark \ref{Ankij=0}, $\weightdist{i}{r-i}=0$. Let us now suppose that $r \geq d = n-k+1$. By Theorem \ref{Ankij}, as $\weightdist{0}{j} = 0$ if $j > 0$,
\begin{displaymath}
\begin{split}
&\sum_{i=0}^{r \wedge k} \weightdist{i}{r-i} = \sum_{i=1}^{r \wedge k} \binom{k}{i} \binom{n-k}{r-i} \cdot \\
&\quad \cdot \sum_{H=0}^{r-n+k-1} (-1)^H \binom{n-k+H-1}{H} (q-1)^{r-n+k-H}.
\end{split}
\end{displaymath}
We calculate $\sum_{i=1}^{r \wedge k} \binom{k}{i} \binom{n-k}{r-i}$. Since $r>n-k$, then $\binom{n-k}{r}=0$ and
\begin{displaymath}
\sum_{i=1}^{r \wedge k} \binom{k}{i} \binom{n-k}{r-i} = \sum_{i=0}^{r \wedge k} \binom{k}{i} \binom{n-k}{r-i}. 
\end{displaymath}
If $r > k$, then $\binom{k}{i} = 0$ for all $k+1 \leq i \leq r$, so, for any $k$ and $r$,
\begin{displaymath}
\sum_{i=0}^{r \wedge k} \binom{k}{i} \binom{n-k}{r-i} = \sum_{i=0}^{r} \binom{k}{i} \binom{n-k}{r-i}.
\end{displaymath}
By Lemma \ref{ChuVandermonde}, we obtain that
\begin{displaymath}
\sum_{i=0}^{r} \binom{k}{i} \binom{n-k}{r-i} = \binom{n}{r},
\end{displaymath}
so $\sum_{i=1}^{r \wedge k} \binom{k}{i} \binom{n-k}{r-i} = \binom{n}{r}$, and hence
\begin{displaymath}
\begin{split}
&\sum_{i=0}^{r \wedge k} \weightdist{i}{r-i} = \\
&= \binom{n}{r} \sum_{H=0}^{r-n+k-1} (-1)^H \binom{n-k+H-1}{H} (q-1)^{r-n+k-H}.
\end{split}
\end{displaymath}
We drop $\binom{n}{r}$ and develop the rest of the previous formula,
\begin{displaymath}
\begin{split}
&\sum_{H=0}^{r-n+k-1} (-1)^H \binom{n-k+H-1}{H} (q-1)^{r-n+k-H} = \\
&= \sum_{H=0}^{r-n+k-1} (-1)^H \binom{n-k+H-1}{H} \cdot \\
&\quad \cdot \sum_{L=0}^{r-n+k-H} (-1)^L \binom{r-n+k-H}{L} q^{r-n+k-H-L} \\
\intertext{reordering the summands}
&= \sum_{H=0}^{r-n+k-1} \sum_{L=0}^{r-n+k-H} (-1)^{H+L} \cdot \\
&\quad \cdot \binom{n-k+H-1}{H} \binom{r-n+k-H}{L} q^{r-n+k-(H+L)} \\
\intertext{computing the coefficient of $q^{r-n+k-j}$, that is, $j=H+L$}
&= \sum_{j=0}^{r-n+k} \sum_{H=0}^{j\wedge(r-n+k-1)} (-1)^j \cdot \\
&\quad \cdot \binom{n-k+H-1}{H} \binom{r-n+k-H}{j-H} q^{r-n+k-j} \\
&= \sum_{j=0}^{r-n+k-1} (-1)^j \cdot \\
&\qquad \cdot \sum_{H=0}^j \binom{n-k+H-1}{H} \binom{r-n+k-H}{j-H} q^{r-n+k-j} + \\
&\quad + \sum_{H=0}^{r-n+k-1} (-1)^{r-n+k} \binom{n-k+H-1}{H},
\end{split}
\end{displaymath}
where we have split the case $j=r-n+k$. 
By Lemma \ref{BinomSumProd} for $\delta = n-k$, $\beta = r$, $\alpha = j$ and $\kappa = H$,
\begin{displaymath}
\sum_{H=0}^j \binom{n-k+H-1}{H} \binom{r-n+k-H}{j-H} = \binom{r}{j},
\end{displaymath}
and this yields
\begin{displaymath}
\begin{split}
&\sum_{i=0}^{r \wedge k} \weightdist{i}{r-i} = \\
&= \binom{n}{r} \sum_{j=0}^{r-n+k-1} (-1)^j \binom{r}{j} q^{r-n+k-j} + \\
&\quad + \binom{n}{r} \sum_{H=0}^{r-n+k-1} (-1)^{r-n+k} \binom{n-k+H-1}{H} \\
&= \binom{n}{r} \sum_{j=0}^{r-d} (-1)^j \binom{r}{j} q^{r-d+1-j} + \\
&\quad + \binom{n}{r} \sum_{H=0}^{r-d} (-1)^{r-d+1} \binom{d-2+H}{H}.
\end{split}
\end{displaymath}
The independent term in $q$ of this expression can be rewritten as
\begin{displaymath}
\begin{split}
&\sum_{H=0}^{r-d} (-1)^{r-d+1} \binom{d-2+H}{H} = \\
&= (-1)^{r-d+1} \cdot \\
&\quad \cdot \sum_{H=0}^{r-d} \binom{d-1+H-1}{H} \binom{(r-1)-(r-d)-H}{(d-1)-H} \\
&= (-1)^{r-d+1} \binom{r-1}{d-1},
\end{split}
\end{displaymath}
where the last equality is given by Lemma \ref{BinomSumProd} with $\beta=r-1$, $\kappa=H$, $\alpha=d-1$ and $\delta=r-d$. By Lemma \ref{altBinom} with $\alpha = r$, $\gamma=j$ and $\kappa=r-d$ we have
\begin{displaymath}
(-1)^{r-d+1} \binom{r-1}{r-d} =  -\sum_{j=0}^{r-d} (-1)^j \binom{r}{j} = \sum_{j=r-d+1}^r (-1)^j \binom{r}{j}
\end{displaymath}
and consequently
\begin{displaymath}
\begin{split}
&\sum_{i=0}^{r \wedge k} \weightdist{i}{r-i} = \\
&= \binom{n}{r} \left[ \sum_{j=0}^{r-d} (-1)^j \binom{r}{j} q^{r-d+1-j} + \sum_{j=r-d+1}^r (-1)^j \binom{r}{j} \right].
\end{split}
\end{displaymath}
This finishes the proof.
\end{IEEEproof}

As a corollary of the above proof, we find a new description of the weight distribution of an MDS code.

\begin{cor}
For any $r\geq 0$, the number of weight $r$ codewords of an MDS code is given by the formula
\begin{displaymath}
\begin{split}
&A_r = \\
&= \binom{n}{r} \sum_{H=0}^{r-n+k-1} (-1)^H \binom{n-k+H-1}{H} (q-1)^{r-n+k-H}.
\end{split}
\end{displaymath}
\end{cor}

Now, we are in the position to describe the information--bit and information--symbol error rate of a false negative, concretely,
\begin{displaymath}
\begin{split}
P^{\text{symbol}}_{\text{FN}} &= \sum_{r=d}^{n}\sum_{i=1}^{r\wedge k} \weightdist{i}{r-i} p_s^{r} q_s^{n-r}\cdot \frac{i}{k} \\
P^{\text{bit}}_{\text{FN}} &= \sum_{r=d}^{n}\sum_{i=1}^{r\wedge k} \weightdist{i}{r-i} p_s^{r} q_s^{n-r}\cdot \frac{i}{k} \cdot p_{b|s}.
\end{split}
\end{displaymath}

\begin{figure}[!ht]
\begin{center}
\includegraphics[scale=0.45]{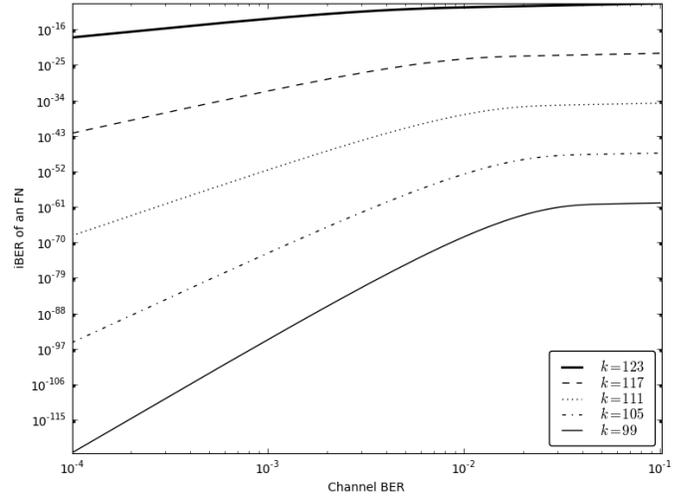}
\caption{iBER of an FN for an MDS code over $\field[128]$ of length 127 and dimension $k$.}\label{graphFN}%
\end{center}
\end{figure}

The iBER of a false negative can be compared for different codes in Figure \ref{graphFN}. Observe that it increases monotonically as the channel BER increases. The iBER of a FN is significantly smaller than the channel BER, at least for dimensions less or equal than $117$.

\section{Decoding failures}\label{s3}

In this section, we shall make use of the calculus of the values $\weightdist{i}{j}$ in order to obtain the information errors of a decoding failure. For simplicity, we may assume that the zero codeword is transmitted. Suppose that we have received an erroneous transmission with $r_1$ non-zero coordinates in the information set and $r_2$ in the redundancy. All along the paper, we shall say that the weight of this error is $r_1+r_2$. If the received word is corrected by the code, then it is at a maximum distance $t$ of a codeword. Obviously, if $0 \leq r_1 + r_2 \leq t$, the word is properly corrected, so we may assume that $t < r_1 + r_2$. In this case, the correction is always wrong and we have a decoding failure. Our aim now is to count these words, highlighting the number of wrong information symbols and the errors belonging to the redundancy, i.e. the words of weight $r_1+r_2$ that decode to a codeword $c$ of weight $c_1+c_2$. The reasoning is as follows: for any codeword $c$ of such weight, we calculate $\wordsaround{(c_1,c_2)}{(r_1,r_2)}$, the number of words of weight $r_1+r_2$ which belong to its sphere of radius $t$. This can be carried out by provoking up to $t$ changes in $c$. Our reasoning is analogous to the one in \cite{DessetMacqVandendorpe2004}. Firstly, there is a minimum number of symbols that must be changed, either to zero or to something non-zero depending on the sign of $r_1-c_1$ and $r_2-c_2$ in order to obtain the correct weight. If $t$ is large enough, we can use the remaining correction capacity to modify an additional number of symbols to zero, and the same number of symbols to a non-zero element of $\field[q]$ in order to keep the weight unchanged. Finally, the remaining possible symbol modifications can be used to change some non-zero symbols into other non-zero symbols, without affecting the weight of the word. Let $\alpha = t - |c_1-r_1| - |c_2-r_2|$ where $|n|$ denotes the absolute value of $n$. We may distinguish four cases:

\paragraph{$r_1 \leq c_1$ and $r_2 \leq c_2$} In the $c_1$ non-zero information coordinates, $c_1-r_1$ of them should be changed to zero. Additionally, we also allow $i_1$ more. In the same way, on the $c_2$ non-zero coordinates of the redundancy, we must change $c_2-r_2$ and $i_2$ of them. Therefore, we have the following number of possibilities:
\begin{displaymath}
\binom{c_1}{c_1-r_1+i_1} \binom{c_2}{c_2-r_2+i_2}.
\end{displaymath}
Now, we should give a non-zero value to $i_1$ coordinates between the $k-c_1$ remaining information symbols, and $i_2$ coordinates between the $n-k-c_2$ remaining redundancy symbols. Thus, the number of possible changes is as follows:
\begin{displaymath}
\binom{k-c_1}{i_1} (q-1)^{i_1} \binom{n-k-c_2}{i_2} (q-1)^{i_2}.
\end{displaymath}
Since the changes cannot exceed $t$, the admissible quantities for $i_1$ and $i_2$ satisfy
\begin{displaymath}
c_1-r_1+i_1+c_2-r_2+i_2+i_1+i_2 \leq t,
\end{displaymath}
and hence
\begin{displaymath}
0 \leq i_1 + i_2 \leq \floor{\frac{\alpha}{2}}.
\end{displaymath}
Finally, we may change some of the remaining non-zero $r_1-i_1$ and $r_2 - i_2$ coordinates to another non-zero symbol. If we change the $j_1$ and $j_2$ coordinates, respectively, we obtain
\begin{displaymath}
\binom{r_1-i_1}{j_1} (q-2)^{j_1} \binom{r_2-i_2}{j_2} (q-2)^{j_2}
\end{displaymath}
changes, where $j_1$ and $j_2$ satisfy
\begin{displaymath}
0 \leq j_1 + j_2 \leq \alpha-2i_1-2i_2.
\end{displaymath}
Therefore, the total number of words is the following:
\begin{displaymath}
\begin{split}
&\wordsaround{(c_1,c_2)}{(r_1,r_2)} = \\
&= \sum_{i_1=0}^{\floor{\frac{\alpha}{2}}} \binom{c_1}{r_1-i_1} \binom{k-c_1}{i_1} \\
&\quad \cdot \sum_{i_2=0}^{\floor{\frac{\alpha}{2}}-i_1} \binom{c_2}{r_2-i_2} \binom{n-k-c_2}{i_2} (q-1)^{i_1+i_2} \\
&\quad \cdot \sum_{j_1=0}^{\alpha-2i_1-2i_2} \binom{r_1-i_1}{j_1} \sum_{j_2=0}^{\alpha-2i_1-2i_2-j_1} \binom{r_2-i_2}{j_2} (q-2)^{j_1+j_2}.
\end{split}
\end{displaymath}
If we denote $I=i_1+i_2$ and $J=j_1+j_2$, by Lemma \ref{prodBinom}, we have
\begin{equation}\label{eq:Nr<<c}
\begin{split}
&\wordsaround{(c_1,c_2)}{(r_1,r_2)} = \\
&= \sum_{I=0}^{\floor{\frac{\alpha}{2}}} \sum_{J=0}^{\alpha-2I} (q-2)^J (q-1)^I \cdot \\
&\quad \cdot \sum_{i=0}^I \binom{c_1}{r_1-i} \binom{c_2}{r_2-I+i} \binom{k-c_1}{i} \binom{n-k-c_2}{I-i} \cdot \\
&\quad \cdot \sum_{j=0}^J \binom{r_1-i}{j} \binom{r_2-I+i}{J-j}\\
&= \sum_{J=0}^\alpha \sum_{I=0}^{\floor{\frac{\alpha-J}{2}}} (q-2)^J (q-1)^I \cdot \\
&\quad\cdot \sum_{j=0}^J \sum_{i=0}^I \binom{k-c_1}{i} \binom{n-k-c_2}{I-i} \cdot \\
&\quad \cdot \binom{c_1}{r_1-i} \binom{c_2}{r_2-I+i} \binom{r_1-i}{j} \binom{r_2-I+i}{J-j} \\
&= \sum_{J=0}^\alpha \sum_{I=0}^{\floor{\frac{\alpha-J}{2}}} (q-2)^J (q-1)^I \cdot \\
&\quad\cdot \sum_{j=0}^J \sum_{i=0}^I \binom{c_1}{j} \binom{c_2}{J-j} \binom{c_1-j}{c_1-r_1+i} \binom{c_2-J+j}{c_2-r_2+I-i} \cdot \\
&\quad \cdot \binom{k-c_1}{i} \binom{n-k-c_2}{I-i}.
\end{split}
\end{equation}

\paragraph{$r_1 \leq c_1$ y $r_2 > c_2$} We proceed as in the above case with the information symbols. So we can make 
\begin{displaymath}
\binom{c_1}{c_1-r_1+i_1} \binom{k-c_1}{i_1} (q-1)^{i_1}
\end{displaymath}
changes in the information set. In the redundancy part, we must give a non-zero symbol to $c_2-r_2+i_2$ coordinates, and, therefore, change $i_2$ of $c_2$ coordinates to zero. In that way, we have
\begin{displaymath}
\binom{c_2}{i_2} \binom{n-k-c_2}{r_2-c_2+i_2} (q-1)^{r_2-c_2+i_2}
\end{displaymath}
possible changes, where
\begin{displaymath}
c_1-r_1+i_1+r_2-c_2+i_2+i_1+i_2 \leq t,
\end{displaymath}
and then
\begin{displaymath}
0 \leq i_1 + i_2 \leq \floor{\frac{\alpha}{2}}.
\end{displaymath}
Finally, changing the value of $j_1$ and $j_2$ of the remaining $r_1-i_1$ and $c_2 - i_2$ non-zero coordinates, we have
\begin{displaymath}
\binom{r_1-i_1}{j_1} (q-2)^{j_1} \binom{c_2-i_2}{j_2} (q-2)^{j_2}
\end{displaymath}
changes, where
\begin{displaymath}
0 \leq j_1 + j_2 \leq \alpha-2i_1-2i_2.
\end{displaymath}
Thus, the total number of words is 
\begin{displaymath}
\begin{split}
&\wordsaround{(c_1,c_2)}{(r_1,r_2)} = \\
&= \sum_{i_1=0}^{\floor{\frac{\alpha}{2}}} \binom{c_1}{r_1-i_1} \binom{k-c_1}{i_1} \cdot \\ 
&\quad \cdot \sum_{i_2=0}^{\floor{\frac{\alpha}{2}}-i_1} \binom{c_2}{i_2} \binom{n-k-c_2}{r_2-c_2+i_2} (q-1)^{i_1+r_2-c_2+i_2} \\
&\quad \cdot \sum_{j_1=0}^{\alpha-2i_1-2i_2} \binom{r_1-i_1}{j_1} \sum_{j_2=0}^{\alpha-2i_1-2i_2-j_1} \binom{c_2-i_2}{j_2} (q-2)^{j_1+j_2}.
\end{split}
\end{displaymath}
Again, we denote $I=i_1+i_2$ and $J=j_1+j_2$, and, by Lemma \ref{prodBinom}, we simplify the expression to
\begin{displaymath}
\begin{split}
&\wordsaround{(c_1,c_2)}{(r_1,r_2)} = \\
&= \sum_{I=0}^{\floor{\frac{\alpha}{2}}} \sum_{J=0}^{\alpha-2I} (q-2)^J (q-1)^{I+r_2-c_2} \cdot \\
&\quad \cdot \sum_{i=0}^I \binom{c_1}{r_1-i} \binom{c_2}{I-i} \binom{k-c_1}{i} \binom{n-k-c_2}{r_2-c_2+I-i} \cdot \\
&\quad \cdot \sum_{j=0}^J \binom{r_1-i}{j} \binom{c_2-I+i}{J-j}\\
&= \sum_{J=0}^\alpha \sum_{I=0}^{\floor{\frac{\alpha-J}{2}}} (q-2)^J (q-1)^{I+r_2-c_2}  \cdot \\
&\quad\cdot \sum_{j=0}^J \sum_{i=0}^I \binom{c_1}{j} \binom{c_2}{J-j} \binom{c_1-j}{c_1-r_1+i} \binom{c_2-J+j}{I-i} \cdot \\
&\quad \cdot \binom{k-c_1}{i} \binom{n-k-c_2}{r_2-c_2+I-i}.
\end{split}
\end{displaymath}
If we change $I' = I+r_2-c_2$ and we take in care that binomial coefficients of negative integers are $0$, we get
\begin{equation}\label{eq:Nr<>c}
\begin{split}
&\wordsaround{(c_1,c_2)}{(r_1,r_2)} = \\
&= \sum_{J=0}^\alpha \sum_{I'=0}^{\floor{\frac{\alpha-J}{2}}+r_2-c_2} (q-2)^J (q-1)^{I'} \cdot \\
&\quad\cdot \sum_{j=0}^J \sum_{i=0}^{I'} \binom{c_1}{j} \binom{c_2}{J-j} \binom{c_1-j}{c_1-r_1+i} \binom{c_2-J+j}{c_2-r_2+I'-i} \cdot \\
&\quad \cdot \binom{k-c_1}{i} \binom{n-k-c_2}{I'-i}.
\end{split}
\end{equation}

\paragraph{$r_1 > c_1$ y $r_2 \leq c_2$} In this case,
\begin{displaymath}
\begin{split}
&\wordsaround{(c_1,c_2)}{(r_1,r_2)} = \\
&= \sum_{J=0}^\alpha \sum_{I=0}^{\floor{\frac{\alpha-J}{2}}} (q-2)^J (q-1)^{I+r_1-c_1} \cdot \\
&\quad\cdot \sum_{j=0}^J \sum_{i=0}^I \binom{c_1}{j} \binom{c_2}{J-j}\binom{c_1-j}{i} \binom{c_2-J+j}{c_2-r_2+I-i} \cdot \\
&\quad \cdot \binom{k-c_1}{r_1-c_1+i} \binom{n-k-c_2}{I-i}
\end{split}
\end{displaymath}
as before. If we make the change of variable $i'=I-i$ we can proceed as in the previous case and we get the formula

\begin{equation}\label{eq:Nr><c}
\begin{split}
&\wordsaround{(c_1,c_2)}{(r_1,r_2)} = \\
&= \sum_{J=0}^\alpha \sum_{I'=0}^{\floor{\frac{\alpha-J}{2}}+r_1-c_1} (q-2)^J (q-1)^{I'} \cdot \\
&\quad\cdot \sum_{j=0}^J \sum_{i''=0}^{I'} \binom{k-c_1}{i''} \binom{n-k-c_2}{I'-i''} \cdot \\
&\quad \cdot \binom{c_1}{j} \binom{c_2}{J-j} \binom{c_1-j}{c_1-r_1+i''} \binom{c_2-J+j}{c_2-r_2+I'-i''}
\end{split}
\end{equation}
where $I'=I+r_1-c_1$ and $i''=I'-i'$.

\paragraph{$r_1>c_1$ y $r_2>c_2$} In this case,
\begin{displaymath}
\begin{split}
&\wordsaround{(c_1,c_2)}{(r_1,r_2)} = \\
&= \sum_{J=0}^\alpha \sum_{I=0}^{\floor{\frac{\alpha-J}{2}}} (q-2)^J (q-1)^{I+r_1-c_1+r_2-c_2} \cdot \\
&\quad\cdot \sum_{j=0}^J \sum_{i=0}^I \binom{c_1}{j} \binom{c_2}{J-j} \binom{c_1-j}{i} \binom{c_2-J+j}{I-i} \cdot \\
&\quad \cdot \binom{k-c_1}{r_1-c_1+i} \binom{n-k-c_2}{r_2-c_2+I-i}.
\end{split}
\end{displaymath}
We proceed as in the two previous cases and we obtain the formula
\begin{equation}\label{eq:Nr>>c}
\begin{split}
&\wordsaround{(c_1,c_2)}{(r_1,r_2)} = \\
&= \sum_{J=0}^\alpha \sum_{I''=0}^{\floor{\frac{\alpha-J}{2}}+r_1-c_1+r_2-c_2} (q-2)^J (q-1)^{I''} \cdot \\
&\quad\cdot \sum_{j=0}^J \sum_{i''=0}^{I''} \binom{k-c_1}{i''} \binom{n-k-c_2}{I''-i''} \cdot \\
&\quad \cdot \binom{c_1}{j} \binom{c_2}{J-j} \binom{c_1-j}{c_1-r_1+i''} \binom{c_2-J+j}{c_2-r_2+I''-i''}
\end{split}
\end{equation}
where $I' = I+r_2-c_2$, $i' = I'-i$, $I''=I'+r_1-c_1$ and $i''=I''-i'$.

\begin{thm}\label{Nrc}
Let $t < r_1 + r_2$. For any codeword $c$ of weight $c_1+c_2$ the number $\wordsaround{(c_1,c_2)}{(r_1,r_2)}$ of words of weight $r_1+r_2$ which belong to its sphere of radius $t$ is 
\begin{displaymath}
\begin{split}
&\wordsaround{(c_1,c_2)}{(r_1,r_2)} = \\
&= \sum_{J=0}^\alpha \sum_{I=0}^{\floor{\frac{\beta-J}{2}}} (q-2)^J (q-1)^I \binom{r_1 + r_2 - I}{J} \cdot \\
&\quad\cdot \sum_{i=0}^I \binom{c_1}{r_1 - i} \binom{c_2}{r_2 - I + i} \binom{k-c_1}{i} \binom{n-k-c_2}{I-i} 
\end{split}
\end{displaymath}
where $\alpha = t - \mid c_1 - r_1 \mid - \mid c_2 - r_2 \mid$ and $\beta = t - c_1 + r_1 - c_2 + r_2$.
\end{thm}

\begin{IEEEproof}
It is a direct consequence of equations \eqref{eq:Nr<<c}, \eqref{eq:Nr<>c}, \eqref{eq:Nr><c} and \eqref{eq:Nr>>c} than
\begin{displaymath}
\begin{split}
&\wordsaround{(c_1,c_2)}{(r_1,r_2)} = \\
&= \sum_{J=0}^\alpha \sum_{I=0}^{\floor{\frac{\beta-J}{2}}} (q-2)^J (q-1)^I \cdot \\
&\quad\cdot \sum_{j=0}^J \sum_{i=0}^I \binom{c_1}{j} \binom{c_2}{J-j} \binom{c_1-j}{c_1-r_1+i} \binom{c_2-J+j}{c_2-r_2+I-i} \cdot \\
&\quad \cdot \binom{k-c_1}{i} \binom{n-k-c_2}{I-i}
\end{split}
\end{displaymath}
By Lemma \ref{prodBinom} 
\begin{gather*}
\binom{c_1}{j} \binom{c_1-j}{c_1-r_1+i} = \binom{c_1}{r_1-i} \binom{r_1-i}{j} \\
\binom{c_2}{J-j} \binom{c_2-J+j}{c_2-r_2+I-i} = \binom{c_2}{r_2-I+i} \binom{r_2-I+i}{J-j},
\end{gather*}
hence
\begin{equation}\label{eq:NtoD}
\begin{split}
&\wordsaround{(c_1,c_2)}{(r_1,r_2)} = \\
&= \sum_{J=0}^\alpha \sum_{I=0}^{\floor{\frac{\beta-J}{2}}} (q-2)^J (q-1)^I \cdot \\
&\quad\cdot \sum_{i=0}^I \binom{c_1}{r_1-i} \binom{c_2}{r_2-I+i} \binom{k-c_1}{i} \binom{n-k-c_2}{I-i} \cdot \\
&\quad \cdot \sum_{j=0}^J \binom{r_1-i}{j} \binom{r_2-I+i}{J-j}
\end{split}
\end{equation}
Lemma \ref{ChuVandermonde} provides the result. 
\end{IEEEproof}

\begin{figure}[!ht]
\begin{center}
\includegraphics[scale=0.45]{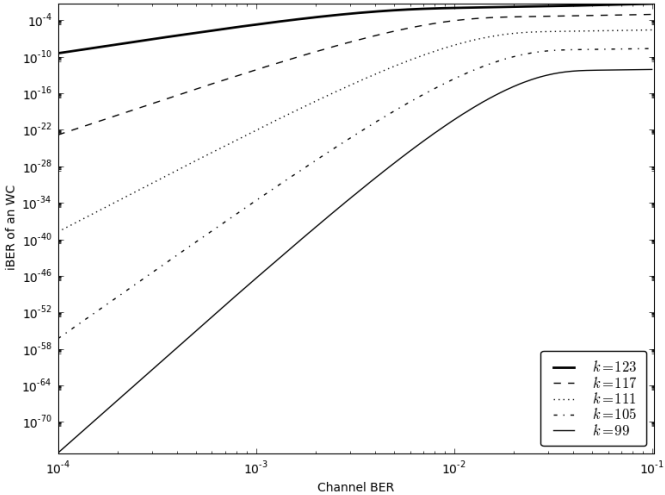}
\caption{iBER of a WC for an MDS code over $\field[128]$ of length 127 and dimension $k$.}\label{graphWC}%
\end{center}
\end{figure}

In order to calculate the iBER in a wrong correction, we follow again the ideology of \cite{DessetMacqVandendorpe2004}. Indeed, the probability of getting a wrong bit is different if it is due to the channel or due to the encoder. In the first case, it is given by the channel BER whilst, in the second case, it is assumed to be $1+\frac{1}{q-1}$. Hence, we should calculate the rate of errors in the information set committed by the decoder, that we shall call $\decoderchanges{(c_1,c_2)}{(r_1,r_2)}$. This calculation is similar to the one for numbers $\wordsaround{(c_1,c_2)}{(r_1,r_2)}$. Concretely, if $r_1 \leq c_1$, for each $i$, $I$, $j$, and $J$, in equation \eqref{eq:NtoD}, the number of changes due to the decoder is $c_1-r_1+i+j$. In order to simplify the resulting expression, note that
\begin{displaymath}
\begin{split}
&\sum_{j=0}^J \binom{r_1-i}{j}\binom{r_2-I+i}{J-j}(c_1-r_1+i+j) = \\
&= (c_1-r_1+i)\binom{r_1+r_2-I}{J} + \\
&\quad + \sum_{j=1}^J \binom{r_1-i}{j}\binom{r_2-I+i}{J-j}j \\
&= (c_1-r_1+i)\binom{r_1+r_2-I}{J} + \\
&\quad + \sum_{j=1}^J \frac{r_1-i}{j}\binom{r_1-i-1}{j-1}\binom{r_2-I+i}{J-j}j \\
&= (c_1-r_1+i)\binom{r_1+r_2-I}{J} + \\
&\quad + (r_1-i)\sum_{\kappa=0}^{J-1} \binom{r_1-i-1}{\kappa}\binom{r_2-I+i}{J-1-\kappa} \\
&= (c_1-r_1+i)\binom{r_1+r_2-I}{J} + \\
&\quad + (r_1-i)\binom{r_1+r_2-I-1}{J-1}
\end{split}
\end{displaymath}
by Lemmas \ref{extBinom} and \ref{ChuVandermonde}, where $\kappa = j-1$. Hence, number $\decoderchanges{(c_1,c_2)}{(r_1,r_2)}$, for the case $r_1\leq c_1$, is derived from equation \eqref{eq:NtoD}
\begin{displaymath}
\begin{split}
&\decoderchanges{(c_1,c_2)}{(r_1,r_2)} = \\
&= \frac{1}{\wordsaround{(c_1,c_2)}{(r_1,r_2)}} \cdot \sum_{J=0}^\alpha \sum_{I=0}^{\floor{\frac{\beta-J}{2}}} (q-2)^J (q-1)^I \cdot \\
&\quad\cdot \sum_{i=0}^I \binom{c_1}{r_1-i} \binom{c_2}{r_2-I+i} \binom{k-c_1}{i} \binom{n-k-c_2}{I-i} \cdot \\
&\quad\cdot \sum_{j=0}^J \binom{r_1-i}{j} \binom{r_2-I+i}{J-j} (c_1-r_1+i+j) \\
&= \frac{1}{\wordsaround{(c_1,c_2)}{(r_1,r_2)}} \cdot \sum_{J=0}^\alpha \sum_{I=0}^{\floor{\frac{\beta-J}{2}}} (q-2)^J (q-1)^I \cdot \\
&\quad\cdot \sum_{i=0}^I \binom{c_1}{r_1 - i} \binom{c_2}{r_2 - I + i} \binom{k-c_1}{i} \binom{n-k-c_2}{I-i} \cdot \\
&\quad\cdot \left( (c_1-r_1+i)\binom{r_1+r_2-I}{J} + \right. \\
&\qquad \left. + (r_1-i)\binom{r_1+r_2-I-1}{J-1} \right),
\end{split}
\end{displaymath}
If $r_1 > c_1$ the number of changes due to the decoder is $i+j$. In this case we check as before that 
\begin{displaymath}
\begin{split}
&\sum_{j=0}^J \binom{r_1-i}{j}\binom{r_2-I+i}{J-j}(i+j) = \\
&= (i)\binom{r_1+r_2-I}{J}+(r_1-i)\binom{r_1+r_2-I-1}{J-1},
\end{split}
\end{displaymath}
so 
\begin{displaymath}
\begin{split}
&\decoderchanges{(c_1,c_2)}{(r_1,r_2)} = \\
&= \frac{1}{\wordsaround{(c_1,c_2)}{(r_1,r_2)}} \cdot \sum_{J=0}^\alpha \sum_{I=0}^{\floor{\frac{\beta-J}{2}}} (q-2)^J (q-1)^I \cdot \\
&\quad\cdot \sum_{i=0}^I \binom{c_1}{r_1 - i} \binom{c_2}{r_2 - I + i} \binom{k-c_1}{i} \binom{n-k-c_2}{I-i} \cdot \\
&\quad\cdot \left( (i)\binom{r_1+r_2-I}{J} + (r_1-i)\binom{r_1+r_2-I-1}{J-1} \right),
\end{split}
\end{displaymath}where $\alpha$ and $\beta$ are as in Theorem \ref{Nrc}. Hence, the information--bit and --symbol error rate of a wrong correction is as follows
. Observe from Figure \ref{graphWC} that the iBER of a WC also increases monotonically, so MDS codes can be said ``proper'' with respect to the iBER, see \cite{Dodunekova1997, KasamiLin1984}.

\begin{displaymath}
\begin{split}
P^{\text{symbol}}_{\text{WC}} &=\sum_{r=d}^{n}\sum_{i=1}^{r\wedge k} \weightdist{i}{r-i} \sum_{r_1=0}^k\sum_{r_2=0}^{n-k} \left (\wordsaround{(i,r-i)}{(r_1,r_2)}-1\right )p_s^{r} q_s^{n-r} \cdot \frac{i}{k} \\
P^{\text{bit}}_{\text{WC}} &=\sum_{r=d}^{n}\sum_{i=1}^{r\wedge k} \weightdist{i}{r-i} \sum_{r_1=0}^k\sum_{r_2=0}^{n-k} \left (\wordsaround{(i,r-i)}{(r_1,r_2)}-1\right )p_s^{r} q_s^{n-r}\cdot \\
&\quad \cdot \frac{1}{k} \left( (i-\decoderchanges{(i,r-i)}{(r_1,r_2)})p_{b|s} + \decoderchanges{(i,r-i)}{(r_1,r_2)}\left(1+\frac{1}{q-1}\right) \right).
\end{split}
\end{displaymath}

\section{False positives}\label{s4}

As we pointed out in the Introduction, there exists the possibility of occurrence of an FP. Up to our knowledge, this has not been treated before in the literature. In this section, we calculate the probability that a PED and an FP occur, finishing our estimation of the iBER of an MDS code. As we noticed above, without loss of generality, we may suppose that the zero word is transmitted and we want to analyze the behaviour of the received word. Our purpose now is to count the words whose weight decomposes as $0+r$, i.e. there are no non-zero information symbols, which are not corrected by the decoder. Obviously, if $0 \leq r \leq t$, the word shall be properly corrected, so we assume that $t+1 \leq r \leq n-k = d-1$. Two disjoint cases can take place: the error is detected but not corrected, producing an FP, or the error is (wrongly) corrected to a codeword. Since the total number of such words is given by
\begin{displaymath}
\sum_{h=t+1}^{n-k} \binom{n-k}{h}(q-1)^h,
\end{displaymath}
it is enough to calculate the words corresponding to one of the two cases. We can make use of the calculations in Section \ref{s3} and give an expression of the words belonging to the second case. Indeed, the number of words of weight $r$ with $t+1\leq r\leq n-k$ that are corrected to a codeword is as follows:
\begin{displaymath}
C(r)=\sum_{s=d}^n\sum_{i=1}^{k\wedge s}\weightdist{i}{s-i}\wordsaround{(i,s-i)}{(0,r)}.
\end{displaymath}
Hence, the number of false positives of weight $r$ is given by
\begin{displaymath}
\mathrm{FP}(r)=\binom{n-k}{r}(q-1)^r-C(r)
\end{displaymath}
and the probability of producing a false positive is given by the following formula
\begin{displaymath}
P_{\mathrm{FP}}=\sum_{r=t+1}^{n-k} FP(r)p_s^rq_s^{n-r}
\end{displaymath}

\begin{figure}[!ht]
\begin{center}
\begin{subfigure}{0.45\textwidth}
\includegraphics[scale=0.4]{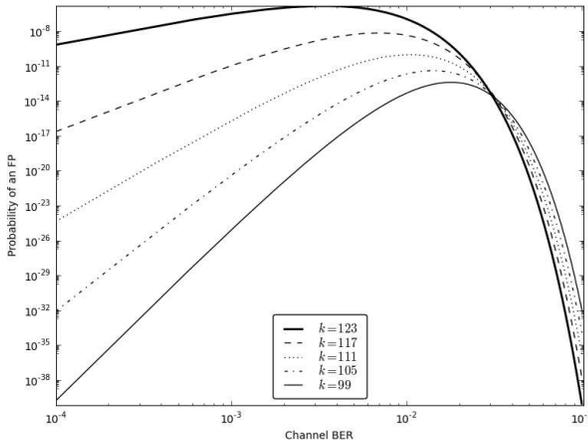}
\caption{Standard range}
\label{graphFP}%
\end{subfigure}
\hfill
\begin{subfigure}{0.45\textwidth}
\includegraphics[scale=0.4]{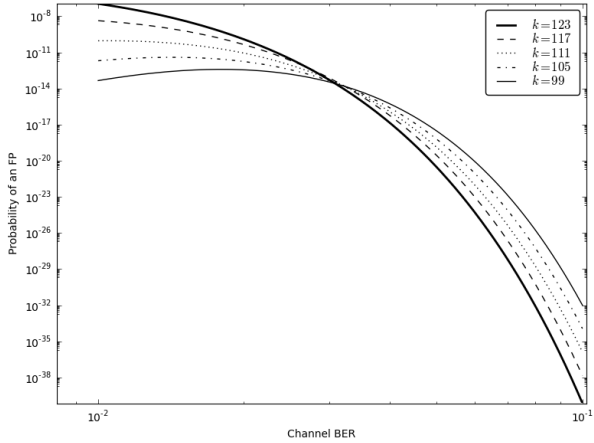}
\caption{Restricted range.}\label{graphFP2}%
\end{subfigure}
\caption{Probability of an FP for an MDS code over $\field[128]$ of length 127 and dimension $k$.}
\end{center}
\end{figure}

It can be observed in Figure \ref{graphFP} that the probability of a FP has a maximum for each code. When the channel BER is high enough this probability increases as the error correction capability of the code increases, see Figure \ref{graphFP2}.

We may now give an estimation of the iBER of a PED. Indeed, when the received word has a weight greater than $t$, the error-correcting capability of the code, three disjoint cases can take place: an undetected error, an FP, or a PED. Hence, for a given weight $i_1+i_2$, the number of words producing a PED is given by
\begin{displaymath}
\begin{split}
\mathrm{PED}(i_1,i_2) &= \binom{k}{i_1}\binom{n-k}{i_2}(q-1)^{i_1+i_2} - \\
&\quad - \mathrm{FP}(i_2) - \sum_{h=d}^{n}\sum_{i=1}^k \weightdist{i}{h-i}\wordsaround{(i,h-i)}{(i_1,i_2)}
\end{split}
\end{displaymath} 
whenever $i_1+i_2>t$ and zero otherwise. Therefore,
\begin{displaymath}
\begin{split}
P^{\text{word}}_{\text{PED}} &= \sum_{i_1=1}^k\sum_{i_2=1}^{n-k} \mathrm{PED}(i_1,i_2) p_s^{i_1+i_2}q_s^{n-i_1-i_2} \\
P^{\text{symbol}}_{\text{PED}} &= \sum_{i_1=1}^k\sum_{i_2=1}^{n-k} \mathrm{PED}(i_1,i_2) p_s^{i_1+i_2}q_s^{n-i_1-i_2} \cdot \frac{i_1}{k} \\
P^{\text{bit}}_{\text{PED}} &=\sum_{i_1=1}^k\sum_{i_2=1}^{n-k} \mathrm{PED}(i_1,i_2) p_s^{i_1+i_2}q_s^{n-i_1-i_2} \cdot \frac{i_1}{k} \cdot p_{b|s}.
\end{split}
\end{displaymath}

\begin{figure}[!ht]
\begin{center}
\begin{subfigure}{0.45\textwidth}
\includegraphics[scale=0.4]{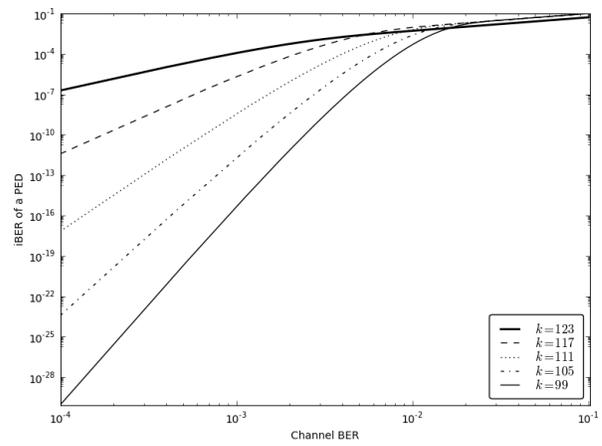}
\caption{Standard range}\label{graphPED}%
\end{subfigure}
\hfill
\begin{subfigure}{0.45\textwidth}
\includegraphics[scale=0.4]{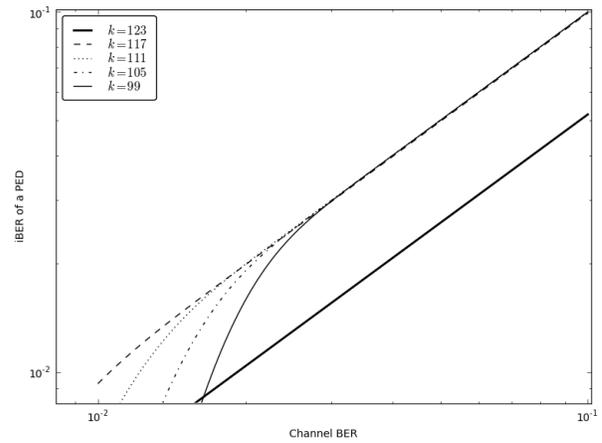}
\caption{Restricted range.}\label{graphPED2}%
\end{subfigure}
\caption{iBER of a PED for an MDS code over $\field[128]$ of length 127 and dimension $k$.}\label{graphPED0}
\end{center}
\end{figure}

The reader may observe from Figures \ref{graphPED0} that, for high channel BER's, the behaviour of the iBER of a PED becomes almost linear. Actually, the curves approximate to the line $y=x$ according the dimension of the code diminishes.

%

\appendices
\section{Some combinatorial identities}

For the convenience of the reader and in order to make the paper self-contained, we add the combinatorial identities that have been referenced all along the paper.

\begin{lem}[Chu--Vandermonde Identity]\label{ChuVandermonde}
\begin{displaymath}
\sum_{\kappa=0}^\rho \binom{\alpha}{\kappa} \binom{\beta}{\rho-\kappa} = \binom{\alpha+\beta}{\rho}.
\end{displaymath}
\end{lem}

\begin{IEEEproof}
See e.g. \cite[page 8, Equation (3)]{Riordan79}.
\end{IEEEproof}

\begin{lem}\label{prodBinom}\label{extBinom}
\begin{displaymath}
\begin{gathered}
\binom{\alpha}{\gamma} \binom{\alpha-\gamma}{\kappa} = \binom{\alpha}{\kappa} \binom{\alpha-\kappa}{\gamma}, \\
\frac{\alpha}{\kappa} \binom{\alpha-1}{\kappa-1} = \binom{\alpha}{\kappa}, \\
\binom{\alpha}{\beta} \binom{\beta}{\gamma} = \binom{\alpha}{\gamma} \binom{\alpha-\gamma}{\beta-\gamma}.
\end{gathered}
\end{displaymath}
\end{lem}

\begin{IEEEproof}
It follows directly from the definition.
\end{IEEEproof}

\begin{lem}\label{altBinom}
\begin{displaymath}
\sum_{\gamma=0}^\kappa (-1)^\gamma \binom{\alpha}{\gamma} = (-1)^\kappa \binom{\alpha-1}{\kappa}.
\end{displaymath}
\end{lem}

\begin{IEEEproof}
An easy induction on $\kappa$ using Pascal's rule. 
\end{IEEEproof}

\begin{lem}\label{BinomSumProd}
\begin{displaymath}
\binom{\beta}{\alpha} = \sum_{\kappa\geq0} \binom{\beta-\delta-\kappa}{\alpha-\kappa} \binom{\delta+\kappa-1}{\kappa}.
\end{displaymath}
\end{lem}

\begin{IEEEproof}
\cite[pg.7, Equation (2)]{Riordan79}
\end{IEEEproof}

\begin{lem}\label{DifFin}
Let $P(x) \in \field{}[x]$ be a polynomial whose degree is less that $\alpha$. Then
\begin{displaymath}
\sum_{\delta=0}^{\alpha} (-1)^{\alpha-\delta} \binom{\alpha}{\delta} P(\delta) = 0
\end{displaymath}
\end{lem}

\begin{IEEEproof}
Direct consequence of the calculus of finite differences, see e.g. \cite[page 91]{vanLintWilson2001}.
\end{IEEEproof}

\section{Proof of Proposition \ref{pwe=Ankij}.} \label{pf:pwe=Ankij}

\begin{IEEEproof}
Let's develop the formula of Theorem \ref{Ankij}: 
\begin{displaymath}
\begin{split}
&\weightdist{i}{j} = \\
&= \binom{k}{i} \binom{n-k}{j} \cdot \\
&\quad \cdot \sum_{H=0}^{i-n+k+j-1} (-1)^H \binom{n-k+H-1}{H} (q-1)^{i-n+k+j-H} \\
&= \binom{k}{i} \binom{n-k}{j} \sum_{H=0}^{i-n+k+j-1} (-1)^H \binom{n-k+H-1}{H} \cdot \\
&\quad \cdot \sum_{\mu=0}^{i-n+k+j-H} \binom{i-n+k+j-H}{\mu} (-1)^\mu q^{i-n+k+j-H-\mu} \\
&= \binom{k}{i} \binom{n-k}{j} \sum_{H=0}^{i-n+k+j-1} \sum_{\mu=0}^{i-n+k+j-H} (-1)^{H+\mu} \cdot \\
&\quad \cdot \binom{n-k+H-1}{H}  \binom{i-n+k+j-H}{\mu} q^{i-n+k+j-H-\mu}.
\end{split}
\end{displaymath}
Let's compute the coefficients of each degree in $q$. Let $\alpha = H + \mu$. The coefficient of $q^0$ is obtained when $\alpha = i-n+k+j$:
\begin{displaymath}
\begin{split}
&\binom{k}{i} \binom{n-k}{j} \sum_{H=0}^{i-n+k+j-1} (-1)^{i-n+k+j} \cdot \\
&\quad \cdot \binom{n-k+H-1}{H} \binom{i-n+k+j-H}{i-n+k+j-H} \\
&= \binom{k}{i} \binom{n-k}{j} (-1)^{i-n+k+j} \binom{i+j-1}{i-n+k+j-1}
\end{split}
\end{displaymath}
where the last equality follows from Lemma \ref{BinomSumProd}. If $0 \leq \alpha \leq i-n+k+j-1$, the coefficient of $q^{i-n+k+j-\alpha}$ is 
\begin{displaymath}
\begin{split}
&\binom{k}{i} \binom{n-k}{j} \cdot \\
&\quad \cdot \sum_{H=0}^{\alpha} (-1)^\alpha \binom{n-k+H-1}{H} \binom{i-n+k+j-H}{\alpha-H} = \\
&= \binom{k}{i} \binom{n-k}{j} (-1)^\alpha \binom{i+j}{\alpha}
\end{split}
\end{displaymath}
where we have applied Lemma \ref{BinomSumProd} again in the equality. Therefore 
\begin{equation}\label{Ankij2}
\begin{split}
&\weightdist{i}{j} = \\
&= \binom{k}{i} \binom{n-k}{j} \left( (-1)^{i-n+k+j} \binom{i+j-1}{i-n+k+j-1} + \right. \\
&\quad \left. + \sum_{\alpha=0}^{i-n+k+j-1} (-1)^\alpha \binom{i+j}{\alpha} q^{i-n+k+j-\alpha} \right).
\end{split}
\end{equation}
We proceed with $A^\mathcal{T}(i,j)$ is a similar way. By \eqref{eq:pwe}
\begin{displaymath}
\begin{split}
&A^\mathcal{T}(i,j) = \\
&= \binom{k}{i} \binom{n-k}{j} \sum_{l=0}^{i} \binom{i}{l} (-1)^{i-l} \cdot \\
&\quad \cdot \sum_{h=n-k+1-l}^{j} \binom{j}{h} (-1)^{j-h} (q^{k-n+h+l} - 1) \\
&= \binom{k}{i} \binom{n-k}{j} \cdot \\
&\quad \cdot \left( \sum_{l=0}^{i} \binom{i}{l} (-1)^{i-l} \sum_{h=n-k+1-l}^{j} \binom{j}{h} (-1)^{j-h} (-1) + \right.\\
&\quad \left. + \sum_{l=0}^{i} \binom{i}{l} (-1)^{i-l} \sum_{h=n-k+1-l}^{j} \binom{j}{h} (-1)^{j-h} q^{k-n+h+l} \right) \\
&\overset{\dag}{=} \binom{k}{i} \binom{n-k}{j} \cdot \\
&\quad \cdot \left( (-1)^{i+j+n-k} \binom{i+j-1}{n-k} + \right. \\
&\quad \left. + \sum_{l=0}^{i} \binom{i}{l} (-1)^{i-l} \sum_{h=n-k+1-l}^{j} \binom{j}{h} (-1)^{j-h} q^{k-n+h+l} \right) \\
&= \binom{k}{i} \binom{n-k}{j} \cdot \\
&\quad \cdot \left( (-1)^{i+j+n-k} \binom{i+j-1}{n-k} + \right. \\
&\quad \left. + \sum_{l=0}^{i} \binom{i}{l} (-1)^{i-l} \sum_{\beta=n-k+1}^{j+l} \binom{j}{\beta-l} (-1)^{j-\beta+l} q^{k-n+\beta} \right) \\
&= \binom{k}{i} \binom{n-k}{j} \left( (-1)^{i+j+n-k} \binom{i+j-1}{n-k} + \right. \\
&\quad \left. + \sum_{\beta=n-k+1}^{j} (-1)^{i+j-\beta} \sum_{l=0}^{i} \binom{i}{l}  \binom{j}{\beta-l} q^{k-n+\beta} + \right. \\
&\quad \left. + \sum_{\beta=j+1}^{j+i} (-1)^{i+j-\beta} \sum_{l=\beta-j}^{i} \binom{i}{l}  \binom{j}{\beta-l} q^{k-n+\beta} \right) \\
&\overset{\ddag}{=} \binom{k}{i} \binom{n-k}{j} \left( (-1)^{i+j+n-k} \binom{i+j-1}{n-k} + \right. \\
&\quad \left. + \sum_{\beta=n-k+1}^{j+i} (-1)^{i+j-\beta} \binom{i+j}{\beta} q^{k-n+\beta} \right)
\end{split}
\end{displaymath}
where $\dag$ and $\ddag$ are consequence of Lemma \ref{ChuVandermonde}. Making $\beta = i+j-\alpha$ we are done.
\end{IEEEproof}

\section*{Acknowledgment}

The authors wish to thank Dr. Pascal Vontobel for pointing out the references \cite{KhamyMcEllice2005} and \cite{Khamy2007} and for its comments of the manuscript. The authors also wish to thank Dr. Javier López for his assistance in performing the graphics.

\providecommand{\bysame}{\leavevmode\hbox to3em{\hrulefill}\thinspace}
\providecommand{\MR}{\relax\ifhmode\unskip\space\fi MR }
\providecommand{\MRhref}[2]{%
  \href{http://www.ams.org/mathscinet-getitem?mr=#1}{#2}
}
\providecommand{\href}[2]{#2}

\end{document}